# Link Prediction in Networks with Nodes Attributes
# by Similarity Propagation

Maosheng Jiang[1], Yonxiang Chen[1], Ling Chen[1,2]

[1]Department of Computer Science, Yangzhou University, Yangzhou, 225127, China

[2]State Key Lab of Novel Software Tech, Nanjing University, Nanjing, 210093, China

## Abstract

The problem of link prediction has attracted considerable recent attention from various domains such as sociology, anthropology, information science, and computer sciences. A link prediction algorithm is proposed based on link similarity score propagation by a random walk in networks with nodes attributes. In the algorithm, each link in the network is assigned a transmission probability according to the similarity of the attributes on the nodes connected by the link. The link similarity score between the nodes are then propagated via the links according to their transmission probability. Our experimental results show that it can obtain higher quality results on the networks with node attributes than other algorithms.

## 1.    Instruction

Many social, biological, and information systems can be naturally described as networks, where vertices represent entities and links denote relations or interactions between the vertices. Social network consists of individuals and their relations such as friendship and partnership. In the field of sociology, there has been a long history of social network analysis which investigates the relations among social entities. In recent years, social network analysis has attracted considerable attention from various business perspectives, such as marketing and business process modeling. In social networks, links among entities may vary dynamically. For example, email communications and cooperative interactions are changing over time.

Link prediction is an important task in social network analysis. It detects the hidden links from the observed part of the network, or predicts the future links given the current structure of the network. Link prediction has several applications including predicting relations among individuals such as friendship and partnership, and predicting their future behavior such as communications and collaborations. In social security network, link prediction is used to identify hidden groups of terrorists or criminals [1]. In the networks of human behavior, link prediction is used to detect and classify the behavior and motion of people [2]. Link prediction also has many applications in some domains outside social networks. In sensor networks, link prediction is used to explore the dynamic temporal properties [3], to ensure information transfer secrecy [4], and to make optimal routing [5].

A problem of increasing interest revolves around node attributes [6-10]. Many real-world networks contain rich categorical node attributes, for example, users in Google+ have profiles with attributes including employer, school, occupation, and places lived. In the *attribute-inference problem*, we aim to populate attribute information for network nodes with missing or incomplete attribute data. This scenario often arises in practice when users in online social networks set their profiles to be publicly invisible or create an account without providing any attribute information. The growing interest in this problem is highlighted by the privacy implications associated with attribute inference as well as the importance of attribute information for applications, including people search, collaborative filtering [11], and user identity resolution [12].

Several methods are proposed for link prediction in networks with node attributes, such as relational learning [13-15], matrix factorization, and alignment [16, 17] based approaches, have been proposed to leverage attribute information for link prediction, but they suffer from scalability issues. More recently, Backstrom and Leskovec [18] presented a Supervised Random Walk (SRW) algorithm for link prediction that combines network structure and edge-attribute information, but this approach does not fully leverage node-attribute information, as it only incorporates node information for neighboring.

In this paper, we propose a link prediction algorithm based on probability propagation in networks with nodes attributes. In the algorithm, each link in the network is assigned a transmission probability according to the similarity of the attributes on the nodes connected by the link. The link similarity score between the nodes are then propagated by a random walk via the links according to their transmission probability. Our experimental results show that it can obtain higher quality results on the networks with node attributes than other algorithms.

## 2.Problem definition

A network can be presented by $G = (V, E)$, where $V$ is the set of nodes, and $E$ is the set of edges. $A = [a_{ij}]_{n*n}$ is the adjacency matrix of the network, where $a_{ij} = 1$ means a link exists between nodes $v_i$ and $v_j$, and $a_{ij} = 0$ otherwise. Each node in the network has its attributes reflecting the nature of the object corresponding to this node. For example, in the network of online community, the nodes represent the individuals in the community and the link of nodes shows the relations between friend, such as colleague, kinsfolk, classmates and so on. For each node, the attributes include the person's age, interest, address, profession etc. The relation of classmate of two nodes reflects that they may have similar attributes such as age, education and address. The relation of colleague of two nodes indicates that they may have similar attributes such as address and profession. In the social network, the attribute information of node is an important factor for link prediction. Using attribute information, we can get more accurate link prediction results then using only topological features of network. The more similar attributes two nodes share, the higher probability they have a relationship and are linked in the network.

Let $m$ be the number of attributes of the nodes, we use a vector $T_i = (t_{i1}, t_{i2}, ..., t_{im})$ to represent the attributes of node $v_i$. Using vectors $T_i$ ($i=1,2,...,n$) as rows, the $n*m$ matrix $T = [t_{ij}]_{n*m}$ is formed as the attribute matrix of the network. For the node pair $v_i$, $v_j \in V$, their similarity of their attributes can be obtained by a similarity measurement on vectors $T_i$ and $T_j$, such as correlation coefficient or cosine similarity. Similarity of vectors $T_i$ and $T_j$ also can be computed based on their distance such as Euclidean distance or cosine distance. Denote the attributes similarity between nodes $v_i$ and $v_j$ as $sim(i,j)$, the $n*n$ matrix $Sim = [sim(i,j)]_{n*n}$ is the attribute similarity matrix of the network.

Our goal is to predict potential links in the network using the topological information represented by adjacent matrix $A$, and the attribute information represented by attribute matrix $T$. The final result of link prediction is represented by a link similarity score matrix $S=[s_{ij}]_{n*n}$, where $s_{ij}$ indicates the probability of the existence of links between nodes $v_i$ and $v_j$. The larger value $s_{ij}$ has, the higher likelihood of the link between nodes $v_i$ and $v_j$, therefore probability matrix $S$ can be more intuitive to reflect the presence of link between the nodes.

## 3. SimRank index

Our algorithm is based on the SimRank [19] which is a link-based similarity measure and builds on the approach of previously existing link-based measures. SimRank is based on both a clear human intuition and a solid theoretical background.

SimRank approach is focused on "object-to-object relationships found in many domains of interest" [19]. It is defined in a self-consistent way, according to the assumption that two nodes are similar if they are connected to similar nodes. Note that the given intuition is recursive by nature.

Given a graph $G(V, E)$ consisting of a set of nodes $V$ and a set of links $E$, we use $\Gamma(v) = \{u \mid u \in V, (u,v) \in E\}$ to denote the set of all neighbors of node $v$. For two notes $v$ and $u$ in the graph, we denote their link similarity as $s_{ij}$. As the base case, any object is considered maximally similar to itself, i.e. having a link similarity score of 1 assigned. We define a link similarity matrix $S=[s_{ij}]_{n*n}$, , and the initial value of element $s_{ij}$ is

$$s_{ij}^0 = \begin{cases} 1 & \text{if } (i,j) \in E \\ 0 & \text{otherwise} \end{cases} \qquad (1)$$

Then, the value of elements in $S$ are updated by an iterative computation. In each iteration, link similarity of each node pair $(x,y)$ is updated by the similarities of the node pairs involving $\Gamma(x)$ and $\Gamma(y)$, such as node pair $(a,b)$ in Figure 1. Suppose notes $x$ and $y$ in Figure 1 are similar, node $x$ connects with $a$ and node $y$ connects with $b$, then $a$ and $b$ will be similar to some extent since two nodes are similar if they are connected to similar nodes. We can transfer the link similarity $s_{xy}$ between $x$ and $y$ to the edge $(a,b)$ via the edges $(x,a)$ and $(y,b)$, so as to affect the link similarity $s_{ab}$ between $a$ and $b$.

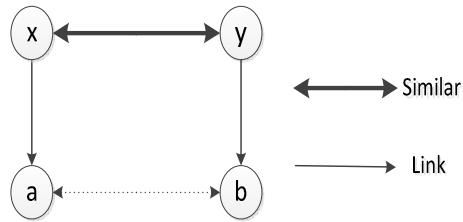

**Figure 1** Link similarity $S(x,y)$ is transmitted to modify $S(a,b)$

Therefore, the iterative formula for calculating $s_{ab}$ is:

$$s_{ab}^{(k+1)} = \frac{c}{|\Gamma(a)| \cdot |\Gamma(b)|} \sum_{x \in \Gamma(a)} \sum_{y \in \Gamma(b)} S_{xy}^{(k+1)} \qquad (2)$$

In formula (2), $c$ is a constant called attenuation coefficient which shows that attenuation degree of link similarity in the transfer process. Such operations of link similarity transferring and modification will be performed repeatedly until the values of link similarity convergence. The final value of $s_{xy}$ is just the link similarity between node pair $(i,j)$, which indicates the probability of the existence of a link between the two nodes.

The SimRank can also be interpreted by the random walk process, that is, $s_{ij}$ measures how soon two random walkers, respectively starting from nodes $i$ and $j$, are expected to meet at a certain node.

## 4. Random Walks model for link prediction in networks with node attributes

A network $G = (V, E)$ with node attributes can be presented by an adjacency matrix $A = [a_{ij}]_{n*n}$ and an attribute matrix $T=[t_{ij}]_{n*m}$. Each row of the attribute matrix $T$ is a vector $T_i = (t_{i1}, t_{i2}, ..., t_{im})$ to represent the attributes of node $v_i$. Each node in the network has its attributes reflecting the nature of the object corresponding to this node, where $m$ is the number of attributes of the nodes.

For link prediction the network with node attributes, we should consider both the topological information of the network represented by the adjacent matrix A, and the attribute information represented by the attribute matrix $T$. Therefore, we first calculate the attribute the attribute similarity between all node pairs in the network. We define the link similarity between two nodes $v_i$ and $v_j$ as the similarity between their attribute vectors $T_i$ and $T_j$. We can use the cosine similarity or correlation coefficient as their similarity. We also can use the similarity based on the distance measure such as Euclidean distance, Hamming distance or Manhattan distance. We use cosine similarity as the attribute similarity measure of the nodes. For two nodes $v_i$ and $v_j$ with attribute vectors $T_i = (t_{i1}, t_{i2}, ..., t_{im})$ and $T_j = (t_{j1}, t_{j2}, ..., t_{jm})$ their attribute similarity is defined as

$$sim(i, j) = \frac{\sum_{k=1}^{m} t_{ik} t_{jk}}{\sqrt{\sum_{k=1}^{m} (t_{ik})^2} \cdot \sqrt{\sum_{k=1}^{m} (t_{jk})^2}} \qquad （3）$$

We define a link similarity matrix $S=[s_{ij}]$, and the initial value of its element $s_{ij}$ as the attribute similarity between nodes $v_i$ and $v_j$, namely $s_{ij} = sim(i,j)$. To transform the link similarity via each edge in the network, we should consider the attribute similarity between the nodes at the end of the edge. Therefore, we assign a transmission probability $P_{ij}$ on each edge $(i,j)$. An edge connecting nodes with large similarity should be assigned higher transmission probability. We define $P_{ij}=sim(i,j)$ on edge $(i,j)$. In Figure 2, let the link similarity between notes $x$ and $y$ be $s_{xy}$. Node $x$ connects with $a$ and node $y$ connects with $b$, and the transmission probabilities on edges $(x,a)$ and $(y,b)$ are $P_{xa}$ and $P_{yb}$ respectively. We can transfer the link similarity $s_{xy}$ between $x$ and $y$ to the edge $(a,b)$ via the edges $(x,a)$ and $(y,b)$, so as to affect the link similarity $s_{ab}$ between $a$ and $b$ with probabilities $P_{xa}$ and $P_{yb}$.

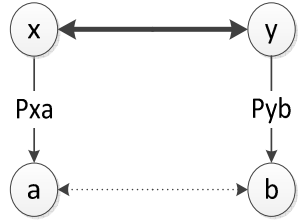

**Figure 2**　Link similarity $s_{xy}$ is transmitted with probabilities $P_{xa}$ and $P_{yb}$

If nodes $x$ and $y$ have high link similarity, and edges edges $(x,a)$ and $(y,b)$ have higher probabilities $P_{xa}$ and $P_{yb}$, then $a$ and $b$ will be similar to some extent since they are closely connected to similar nodes.

Therefore, the iterative formula for calculating $s_{ab}$ is:

$$S_{ab}^{(k+1)} = \begin{cases} \dfrac{c}{d_b.P_a + d_a.P_b} \displaystyle\sum_{x \in \Gamma(a)} \sum_{y \in \Gamma(b)} (P_{xa} + P_{yb}).S_{xy}^{(k)} & a \neq b \\ 1 & a = b \end{cases} \quad (4)$$

Here，$c$ is the attenuation coefficient， $\Gamma(x) = \{v|v \in V, (x,v) \in E\}$ is a collection of neighbors of node x, and $P_x = \displaystyle\sum_{v \in \Gamma(x)} sim(x,v)$ is the summation of similarities between node $x$ and all its neighbors.

The initial value of $S_{xy}$ is defined as follows:

$$S_{xy}^{(0)} = \begin{cases} 1 & x = y \\ 0 & x \neq y \end{cases} \quad (5)$$

The method can also be interpreted as the random walk process as follows. Suppose two walkers respectively starting from nodes *i* and *j*, randomly walk in the network. They pass though edge (*a,b*) at the transmission probability $P_{ab}$. Then $s_{ij}$ measures how soon the walkers are expected to meet at a certain node in the network.

The algorithm for link prediction in networks with node attributes is as follows.

**Algorithm**    *RandWalk*
**Input:** *A*=[$a_{ij}$]: adjacent matrix of the network;
      *T*=[$t_{ij}$] : attribute matrix of the network;
      *c*: attenuation coefficient;
**Output:** *S*=[$s_{ij}$]: similarity matrix between nodes;
**Begin**
  (1)  **For**    *i*=1 **to** *n* **do**
        $P_i$=0;
       **For**    *j*=1 **to** *n* **do**
          Calculate the attribute similarity *sim(i,j)* between nodes $v_i$ and $v_j$ according to (3);
          $s_{ij}$ =*sim(i,j)*; //set the initial value of link similarity between nodes $v_i$ and $v_j$;
          $P_{ij}$=*sim(i,j)*\* $a_{ij}$ // set the transmission probability on edge (*i,j*);
          $P_i$= $P_i$+ $P_{ij}$;
       **Endfor** *j*
     **Endfor** *i*;
  (2)   **While** not convergence **do**
      **For**    *i*=1 **to** *n* **do**
        **For**    *j*=1 **to** *n* **do**
           Update $s_{ij}$ according to (4)
        **Endfor** *j*
      **Endfor** *i*;
     **Endwhile**;
**End**

In step (1), the algorithm first calculates the attribute similarity *sim(i,j)* between all node pairs and the transmission probability matrix *P*=[$P_{ij}$]. In this step, the algorithm also initializes the link similarity matrix *S*=[$s_{ij}$]. In step (2), the algorithm simulates the random walk to update the similarity matrix *S* iteratively until the values of *S* elements converge. Then the final value of $s_{ij}$ is just the link similarity between nodes $v_i$ and $v_j$ which indicates the probability of occurrence of the link between the two nodes.

## 5.  Correctness of the algorithm

In this section, we prove the convergence of the iterative procedure in algorithm *RandWalk,*

and show the existence and uniqueness of the simultaneous solution of equation (4).

**Theorem 1** For every *x, y*, the sequence $\{S_{xy}^{(k)}\}$ defined by (4) converges to a limit $S_{xy}$ by the iterative procedure in algorithm *RandWalk*.

**Proof:** We first prove that

## 6.Experimental results

### 6.1 Experiment Setup

To evaluate our proposed algorithm for link prediction in network with node attributes, we test it by a series of experiments on several data sets of networks. All the experiments are performed on a Pentium IV computer running Windows XP, with 1.7G memory, and using VC++ 6.0.

We use AUC (Area Under Curve) scores to evaluate the quality of the results by the algorithms tested. After the algorithms calculating and ranking the similarities of the all the node pairs, which represent all the existent and the nonexistent links, the AUC value can be interpreted as the probability that a randomly chosen existing link is given a higher score than a randomly chosen non-existing link. At each time we randomly pick an existing link and a non-existing link to compare their scores, if among *n* independent comparisons, there are *n'* times the existing link having a higher score and *n''* times they have the same score, the AUC value is

$$AUC=(n'+0.5n'')/n \qquad (6)$$

In general, a larger AUC value indicates a higher performance, hence, AUC value of the perfect result is 1.0, while AUC of the result by a random predictor is 0.5.

### 5.2 Test data

In our experiments, we test our method on eight representative data sets of networks drawn from Digital Bibliography Library Project which is a computer science bibliography website including: ACM Computer Science Conference (ACM), Computers, Environment and Urban Systems(CEUS), International Conference on Information and Communication Security (ICICS),

International Journal of Computer Graphics & Animation (IJCGA), International Journal of Network Security (IJNS), Journal of Computer-Mediated Communication (JCMC), Mathematical Structures in Computer Science(MSCS) and Natural Language to Data Bases (NLDB).

We extract two matrixes from each network: the adjacency matrix and the attribute matrix which indicates the attribute of each node. In each network, the nodes represent the authors, attributes present the key words in the title of papers. Then the datasets present the links between authors and their papers.

The topological features of these datasets are as shown in Table 1. In the table, $N$ and $M$ are the total numbers of nodes and links, respectively. $Att$ is the number of attributes, while $NUMC$ is the number of the connected components in the network and the size of the largest one. For example, 16/688 means that this network has 688 connected components and the largest one contains 16 nodes. $e$ is the efficiency of the network, $C$ and $r$ are clustering coefficient and assortative coefficient respectively. $K$ is the average degree of the network.

**Table 1**    Topological features of the networks tested

| Networks | $N$ | $M$ | $Att$ | $NUM_C$ | $e$ | $C$ | $r$ | $K$ |
|----------|-----|-----|-------|---------|-----|-----|-----|-----|
| ACM | 1465 | 1209 | 1960 | 16/688 | 0.0014 | 0.3621 | 0.5570 | 1.6505 |
| CEUS | 1047 | 1543 | 1293 | 40/284 | 0.0041 | 0.7092 | 0.4789 | 2.9475 |
| ICICS | 888 | 1398 | 1066 | 187/208 | 0.0139 | 0.7484 | 0.2726 | 3.1486 |
| IJCGA | 940 | 1699 | 1290 | 533/160 | 0.0748 | 0.5955 | 0.1469 | 3.6149 |
| IJNS | 1059 | 1305 | 1263 | 58/309 | 0.0039 | 0.6394 | 0.1733 | 2.4646 |
| JCMC | 1198 | 1477 | 1648 | 39/463 | 0.0032 | 0.5138 | 0.6972 | 2.4658 |
| MSCS | 870 | 825 | 1341 | 160/319 | 0.0080 | 0.4282 | 0.2658 | 1.8966 |
| NLDB | 847 | 1211 | 1041 | 96/225 | 0.0072 | 0.7130 | 0.2112 | 2.8595 |

### 5.3   The results

In our experiment, we test the accuracy of our algorithm *RandWalk*, and compare the AUC of the results with other ten link prediction algorithms *CN, Salton, Jaccard, Sorenson, HPI, HDI ,LHN-I, PA, LP* and *Kaze*. The test results are shown in Table 2.

**Table 2**    Comparison of algorithms' accuracy quantified by AUC

|          | ACM    | CEUS   | ICICS  | IJCGA  | IJNS   | JCMC   | MSCS   | NLDB   |
|----------|--------|--------|--------|--------|--------|--------|--------|--------|
| RandWalk | **0.8280** | 0.9276 | **0.9421** | **0.9201** | **0.9017** | **0.8937** | **0.8842** | **0.9298** |
| CN       | 0.8222 | 0.9319 | 0.9312 | 0.9116 | 0.8964 | 0.8781 | 0.8186 | 0.9091 |
| Salton   | 0.8222 | 0.9321 | 0.9316 | 0.9119 | 0.8966 | 0.8782 | 0.8187 | 0.9093 |
| Jaccard  | 0.4075 | 0.2894 | 0.3602 | 0.5551 | 0.3164 | 0.4737 | 0.4775 | 0.3398 |
| Sorenson | 0.8222 | 0.9321 | 0.9388 | 0.9115 | 0.8965 | 0.8782 | 0.8187 | 0.9093 |
| HPI      | 0.8222 | 0.932  | 0.9388 | 0.9119 | 0.8966 | 0.8781 | 0.8186 | 0.9093 |
| HDI      | 0.8222 | 0.9321 | 0.9387 | 0.9112 | 0.8965 | 0.8781 | 0.8186 | 0.9092 |
| LHN-I    | 0.8222 | 0.932  | 0.9385 | 0.9106 | 0.8964 | 0.878  | 0.8186 | 0.9091 |
| PA       | 0.5440 | 0.514  | 0.5481 | 0.6191 | 0.4676 | 0.6085 | 0.5611 | 0.4909 |
| LP       | 0.8219 | **0.9377** | 0.9408 | 0.907  | 0.8994 | 0.8777 | 0.8175 | 0.916  |
| Kaze     | 0.8219 | **0.9377** | 0.9388 | 0.8847 | 0.8992 | 0.8775 | 0.8139 | 0.9153 |

We can observe from Table 2 that our method has the best performance in all datasets except in CEUS among the 11 algorithms. Although method *Katz* has a deep consideration on global information of common paths, our algorithm *RandWalk* gets even better results in 7 data sets. On datasets of CEUS, AUC value of algorithm *RandWalk* has no significant difference from that of Katz. The reason for our algorithm *RandWalk* getting higher quality results is that it integrates topological and attribute information. It precisely reflects the occurrence probabilities for the given network with node attributes, and eliminates the influence of bias data. This demonstrates that our proposed algorithm *RandWalk* can increase the quality of link prediction in networks with node attributes.

## 7. Conclusions

In this paper, we investigate the problem of link prediction in networks with nodes attributes. A link prediction algorithm *RandWalk* is proposed based on probability propagation in networks with nodes attributes. In the algorithm, each pair of nodes is assigned a transmission probability according to the similarity of the attributes on the pair of nodes. The link similarity score between the nodes are then propagated by the random walk through the links according to their transmission probability. Our experimental results show that algorithm *RandWalk* can obtain higher quality results on the networks with node attributes than other algorithms.

## Acknowledgements

This research was supported in part by the Chinese National Natural Science Foundation under grant Nos. 61379066, 61070047, 61379064，61472344, Natural Science Foundation of Jiangsu Province under contracts BK20130452, BK2012672，BK2012128, and Natural Science Foundation of Education Department of Jiangsu Province under contract 12KJB520019, 13KJB520026, 09KJB20013.